\documentclass[aps,prl,10pt,a4paper,twocolumn,notitlepage,showpacs,showkeys,nosuperscriptaddress]{revtex4-1}

\usepackage{times}
\usepackage{graphicx}
\usepackage{amssymb}
\usepackage{amsmath}
\usepackage[unicode=true, pdfusetitle,
 bookmarks=true,bookmarksnumbered=false,bookmarksopen=false,
 breaklinks=false,pdfborder={0 0 1},backref=false,colorlinks=true]
 {hyperref}

\begin{document}

\title{Modulation of localized solutions for the Schr\"odinger equation with
logarithm nonlinearity}

\author{L. Cala\c{c}a$^1$, A. T. Avelar$^1$, D. Bazeia$^{2,3}$, and W. B. Cardoso$^{1,}$\footnote{Correspondent author.\\ Electronic address: \textit{wesleybcardoso@gmail.com}\\ Tel./Fax.: +55 62 3521-1014}}
\address{$^1$Instituto de F\'{i}sica, Universidade Federal de Goi\'as, 74.001-970,
Goi\^ania, Goi\'as, Brazil\\ $^2$Instituto de F\'{i}sica, Universidade de S\~ao Paulo, 05314-970, S\~ao
Paulo SP, Brazil\\ $^3$Departamento de F\'{i}sica, Universidade Federal da Para\'{i}ba,
58051-970, Jo\~ao Pessoa, Para\'{i}ba, Brazil}

\begin{abstract}
We investigate the presence of localized analytical solutions of the
Schr\"odinger equation with logarithm nonlinearity. After including
inhomogeneities in the linear and nonlinear coefficients, we use similarity
transformation to convert the nonautonomous nonlinear equation into
an autonomous one, which we solve analytically. In particular, we
study stability of the analytical solutions numerically. 
\end{abstract}


\maketitle

\emph{Introduction} - In 1976, Bialynicki-Birula and Mycielsk \cite{Birula76AN}
proposed the logarithmic Schr\"odinger equation (LSE). The aim was to
obtain a nonlinear equation that could be used to quantify departures
from the strictly linear regime, preserving in any number of dimensions
some fundamental aspects of quantum mechanics such as separability
and additivity of total energy of noninteracting subsystems. Although
the LSE possesses very nice properties such as analytic solutions
given by stable Gaussian wave packets in the absence of external forces,
the realization of detailed experiments with ions trap \cite{Bollinger90}
established stringent upper limits on the nonlinear terms in the Schr\"odinger
equation, making the LSE not a general formalism to describe nonlinear
interactions.

Despite losing generality, the LSE has been employed to model nonlinear
behavior in several distinct scenarios in physics and in other areas
of nonlinear science. To be more specific, the LSE appears, for instance,
in dissipative systems \cite{Hernandez80}, in nuclear physics \cite{Hefter85},
in optics \cite{Krolikowski00,Buljan03}, capillary fluids \cite{DeMartinooPROC03},
and even in magma transport \cite{DeMartinoEPL03}. In addition,
some important mathematical contributions have been appeared, namely,
the existence of stable and localized nonspreading Gaussian shapes
\cite{Cazenave83}, dispersion and asymptotic stability features
\cite{Cazenave80}, the existence of unique global mild solution
\cite{Guerrero10}, the obtention of stationary solutions via Lie
symmetry approach \cite{Khalique10}, and the study of optical solitons
with log-law nonlinearity with constant coefficients \cite{BiswasCNSNS10,BiswasOPL10,BiswasJIMT10}.

In the above mentioned applications, one usually focuses on LSE presenting
constant nonlinear coefficient, i.e., without spatial and/or temporal
modulations. However, in a more interesting scenario
the nonlinear parameter that characterizes the physical systems may
depend on space, leading to solutions that can be modulated in space.
The presence of the explicit spatial dependence of the nonlinear term
in the LSE opens interesting perspectives not only from the theoretical
point of view, for investigation of nonuniform nonlinear equations,
but also from the experimental view, for the study of the physical
properties of the systems. {In optical mediums, the modulation of the 
nonlinearity can be achieved in different ways \cite{KartashovRMP11}. 
As an example, in photorefractive media, such as LiNbO$_3$, nonuniform 
doping with Cu or Fe may considerably enhance (modulating) the local nonlinearity 
(as shown in the review paper \cite{HukriedeJPD03}). In this sense, in Ref. \cite{BiswasAMC10} 
the authors have studied the existence of exact 1-soliton solution to 
the nonlinear Schr\"odinger's equation with log law nonlinearity in presence 
of time-dependent perturbations.} Motivated by this, in the present work
we investigate explicit solitonic solutions to the nonuniform LSE.
To achieve this goal, we take advantage of recent works on analytical
solitonic solutions for the cubic \cite{BelmontePRL08}, cubic-quintic
\cite{AvelarPRE09}, quintic \cite{BelmontePLA09}, and coupled
\cite{CardosoPRE12} nonlinear Schr\"odinger equations with space-
and time-dependent coefficients. Analytical breather solutions can
also be constructed for nonuniform nonlinear Schr\"odinger equation
and it has been obtained in \cite{Cardoso1PLA10}.

%%%%%%%%%%%%%%
\emph{Theoretical model} - We consider the LSE given by 
\begin{equation}
i\psi_{z}=-\psi_{xx}+V\psi+g\psi\log|\psi|^{2},\label{LNSE}
\end{equation}
 where $\psi=\psi(x,z)$ with $\psi_{z}\equiv\partial\psi/\partial z$
and $\psi_{xx}\equiv\partial^{2}\psi/\partial x^{2}$. $V=V(x,z)$ 
and $g=g(z)$ are the linear and nonlinear coefficients, respectively. 
To solve (\ref{LNSE}) we use the similarity transformation, taking
the following \emph{ansatz} 
\begin{equation}
\psi=\rho(z)e^{i\eta(x,z)}\Phi[\zeta(x,z),\tau(z)].\label{anzats}
\end{equation}
 Replacing this into Eq.~(\ref{LNSE}) one gets 
\begin{equation}
i\Phi_{\tau}=-\Phi_{\zeta\zeta}+G\Phi\log|\Phi|^{2},\label{LNSE-2}
\end{equation}
 where $G$ is a constant and with the specific forms for the linear
and nonlinear coefficients 
\begin{eqnarray}
V & = & -\eta_{z}-\eta_{x}^{2}-2g\log\rho,\label{pot}\\
g & = & G\;\zeta_{x}^{2},\label{nlin}
\end{eqnarray}
 respectively, plus the following conditional equations 
\begin{eqnarray}
(\rho^{2})_{z}+2\rho^{2}\eta_{xx} & = & 0,\label{c1}\\
\zeta_{z}+2\eta_{x}\zeta_{x} & = & 0,\label{c2}\\
\zeta_{xx} & = & 0,\label{c3}\\
\tau_{z} & = & \zeta_{x}^{2}.\label{c4}
\end{eqnarray}

We see from Eq.~(\ref{c3}) that $\zeta=\alpha(z)x+\beta(z)$. Thus,
using Eq.~(\ref{c2}) we obtain 
\begin{equation}
\eta=-\frac{\alpha_{z}}{4\alpha}x^{2}-\frac{\beta_{z}}{2\alpha}x+\gamma(z),\label{eta}
\end{equation}
 where the function $\gamma(z)$ was introduced after an integration
on the $x$ coordinate. Now, replacing Eq.~(\ref{eta}) into (\ref{c1})
we conclude that
\begin{equation}
\rho = \sqrt{\alpha}.
\label{rho}
\end{equation}
Consequently, from Eq.~(\ref{c4}) we get $\tau=\int\alpha^{2}\mathrm{dz}$.

The above results can be used to rewrite the linear and nonlinear
coefficients in Eqs.~(\ref{pot}) and (\ref{nlin})) in the respective
forms 
\begin{equation}
V=\delta_{1}(z)x^{2}+\delta_{2}(z)x+\delta_{3}(z),\label{newpot}
\end{equation}
 and 
\begin{equation}
g=G\alpha^{2},\label{newnlin}
\end{equation}
 where 
\begin{eqnarray}
\delta_{1} & = & \frac{\alpha_{zz}}{4\alpha}-\frac{\alpha_{z}^{2}}{2\alpha^{2}},\label{d1}\\
\delta_{2} & = & \frac{\beta_{zz}}{2\alpha}-\frac{\alpha_{z}\beta_{z}}{\alpha^{2}},\label{d2}\\
\delta_{3} & = & -\gamma_{z}-\frac{\beta_{z}^{2}}{4\alpha^{2}}-G\alpha^{2}\log\alpha.\label{d3}
\end{eqnarray}

Now, in order to write an explicit solution for the above Eq.~(\ref{LNSE-2})
we consider $\Phi=\phi(\zeta)e^{-i\epsilon\tau}$; this requires that
$\phi$ has to have the form 
\begin{equation}
\phi=\exp\left[\frac{\epsilon+G(1+G\zeta^{2})}{2G}\right],\label{sol}
\end{equation}
 which ends the formal calculations. {We stress that to obtain localized solutions (with a Gaussian shape) it is necessary a self-focusing medium (negative nonlinear coefficient), since we are considering the group velocity dispersion as negative.}

%%%%%%%%%%%%%%%
\emph{Analytical results} - Let us now study specific examples of
modulation of localized solution (\ref{sol}) in the above model.
To do this, we consider distinct values of modulation through the
appropriate choice of $\alpha$, $\beta$, and $\gamma$.

\begin{figure}[tb]
\centering 
\includegraphics[width=0.48\columnwidth]{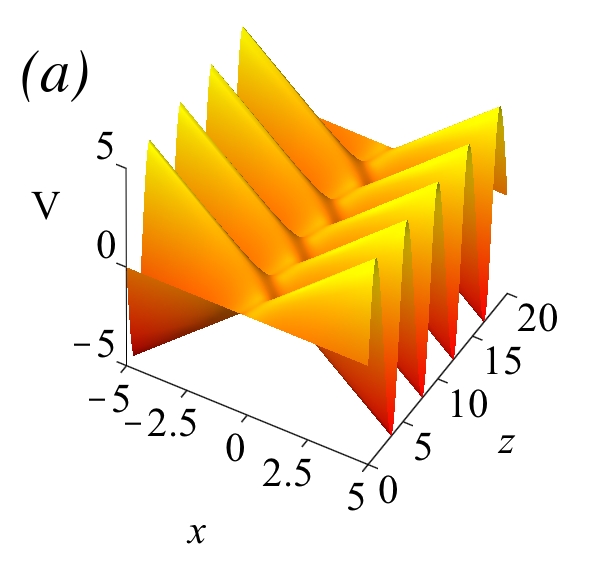}
\includegraphics[width=0.48\columnwidth]{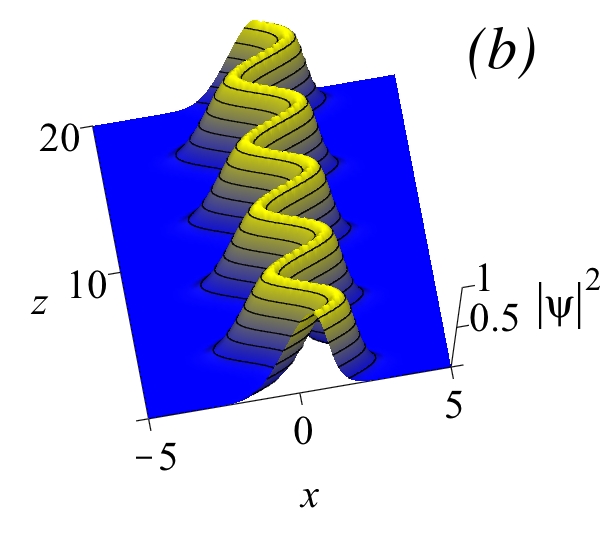}
\caption{(Color online) (a) Linear coefficient (potential) and (b) modulated
solution for the case 1. We have used $G=-1$, $\epsilon=-G$, and
$\omega=\sqrt{2}$.}
\label{FC1} 
\end{figure}

%%%%%%%%%%
\emph{Case \#1} - First we take $\alpha=1$, $\beta=-\sin(\omega z)$,
and $\gamma$ with a specific choice such that $\delta_{3}=0$. Thus,
we have $\delta_{1}=0$ and $\delta_{2}=\omega^{2}\sin(\omega z)/2$.
Here the linear coefficient (\ref{newpot}) assumes a linear behavior
in $x$, with a periodic modulation in the $z$-direction while the
nonlinear coefficient takes a constant value: 
\begin{equation}
V=\frac{\omega^{2}}{2}\sin(\omega z)x\qquad\mathrm{and}\qquad g=G.\label{case1}
\end{equation}
 Note that in this case $\zeta=x-\sin(\omega z)$ and $\tau=z$. Also,
the amplitude and phase of the ansatz (\ref{anzats}) are given by
$\rho=1$ and 
\begin{equation}
\eta=\frac{\omega}{4}\left[2\, x-\omega\cos\left(\omega\, z\right)\right]\cos\left(\omega\, z\right),
\end{equation}
 respectively. In Fig.\ref{FC1} we display the behavior of the linear
coefficient (potential) as well as the field intensity $|\psi|^{2}$,
considering $G=-1$ (self-focusing nonlinearity), $\epsilon=-G$,
and $\omega=\sqrt{2}$. The potential assumes a zigzag behavior that
modules the solution with an oscillatory pattern.

%%%%%%%%%%%%%%%%%%%%
%%%%%%%%%%%%%%%

\begin{figure}[tb]
 \centering 
\includegraphics[width=0.48\columnwidth]{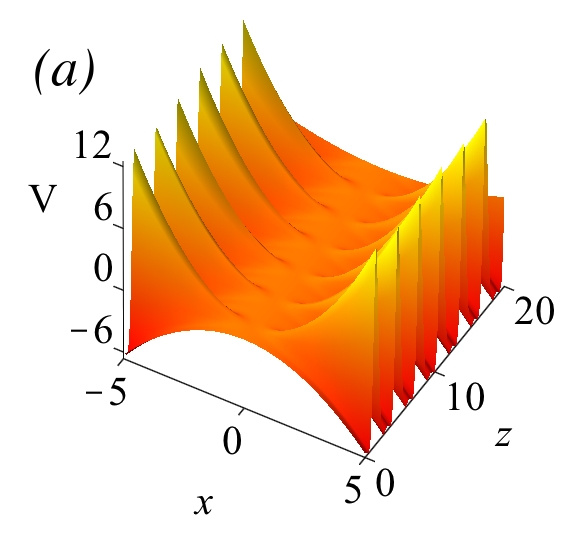}
\includegraphics[width=0.48\columnwidth]{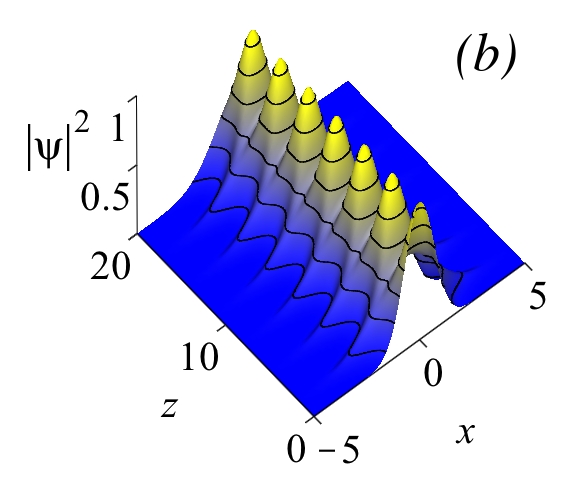} 
\caption{(Color online) (a) Linear coefficient (potential) and (b) modulated
solution for the case 2. We have used $G=-1$ and $\omega=1$.}
\label{FC2} 
\end{figure}

%%%%%%%%%%%%%%%%%%%%%
\emph{Case \#2} - Next, we assume a nonlinear coefficient with an
oscillatory amplitude. As an example, we use $\alpha=\left[1+\cos^{2}(\omega z)\right]/2$,
$\beta=0$, and $\gamma$ with a specific choice such that $\delta_{3}=0$.
In this case, one gets 
\begin{equation}
\zeta=\left[0.5+0.5\cos^{2}(\omega z)\right]x
\end{equation}
 and 
\begin{equation}
\tau=\left\{ \left[11\cos\left(\omega z\right)+2\left(\cos\left(\omega z\right)\right)\right]\sin\left(\omega z\right)+19\omega z\right\} /32\omega.
\end{equation}
 Additionally, we have 
\begin{eqnarray}
V & = & \frac{1-5\cos^{2}\left(\omega z\right)+2\cos^{4}\left(\omega z\right)}{2\left[1+\cos^{2}\left(\omega z\right)\right]^{2}}\omega^{2}x^{2},\label{pot_case2}\\
g & = & \frac{G}{4}\left[1+\cos^{2}\left(\omega z\right)\right]^{2}.\label{nlin_case2}
\end{eqnarray}
 Also, the amplitude and phase of the solution are given by 
\begin{equation}
\rho=\sqrt{1+\cos^{2}(\omega z)}/\sqrt{2},
\end{equation}
 and 
\begin{equation}
\eta=\frac{\omega\cos\left(\omega z\right)\sin\left(\omega z\right)}{2[1+\cos^{2}\left(\omega z\right)]}x^{2}+\gamma,
\end{equation}
 respectively, where 
\begin{equation}
\gamma=-\frac{1}{4}G[1+\cos^{2}(\omega z)]^{2}\ln[(1+\cos^{2}(\omega z))/2].
\end{equation}
 This allows us to find a new analytical solution for $\psi$. In
Fig. \ref{FC2} we show the profile of $|\psi|^{2}$ considering $G=-1$
(self-focusing nonlinearity), $\epsilon=-G$, and $\omega=1$. The
potential assumes a flying-bird behavior that modulates the solution
with a breathing pattern.

%%%%%%%%%%%%%%%%
%%%%%%%%%%%%%%

%%%%%%%%%%
\emph{Case \#3} - Another example can be introduced, after considering
a linear potential with a combination of linear and quadratic terms
in $x$, plus a periodic modulation in the $z$ coordinate. To this
end, we can take 
\begin{equation}
\alpha=\frac{1}{2}\left[1+\cos^{2}(\omega_{1}z)\right],
\end{equation}
 
\begin{equation}
\beta=-2\sin(\omega_{2}z),
\end{equation}
 and $\gamma$ with a specific choice such that $\delta_{3}=0$. These
choices allow us to write 
\begin{equation}
\zeta=\frac{1}{2}\left[1+\cos^{2}(\omega_{1}z)\right]x-2\sin(\omega_{2}z),
\end{equation}
 
\begin{equation}
\tau=\frac{1}{32\omega_{1}}\left\{ \left[11\cos\left(\omega_{1}z\right)+2\cos^{3}\left(\omega_{1}z\right)\right]\sin\left(\omega_{1}z\right)+19\omega_{1}z\right\} ,
\end{equation}
 
\begin{equation}
\delta_{1}=\frac{\omega_{1}^{2}\left[1-5\cos^{2}\left(\omega_{1}z\right)+2\cos^{4}\left(\omega_{1}z\right)\right]}{2\left[1+\cos^{2}\left(\omega_{1}z\right)\right]^{2}},
\end{equation}
 and 
\begin{eqnarray}
\delta_{2} & = & \frac{1}{\left[1+\cos^{2}\left(\omega_{1}z\right)\right]^{2}}\big\{2\omega_{2}\left[\omega_{2}+\omega_{2}\cos^{2}\left(\omega_{1}z\right)\right]\sin\left(\omega_{2}z\right)\nonumber \\
 & - & 8\omega_{2}\cos\left(\omega_{1}z\right)\omega_{1}\cos\left(\omega_{2}z\right)\sin\left(\omega_{1}z\right)\big\}.
\end{eqnarray}
 Thus, we get the expected form $V=\delta_{1}x^{2}+\delta_{2}x$ and
\begin{equation}
g=\frac{G}{4}\left[1+\cos^{2}\left(\omega_{1}z\right)\right]^{2}.
\end{equation}
 Also, the amplitude and phase can be written in the form 
\begin{equation}
\rho=\sqrt{\frac{1}{2}[1+\cos^{2}(\omega_{1}z)]},
\end{equation}
 and 
\begin{equation}
\eta=\frac{\omega_{1}\cos\left(\omega_{1}z\right)\sin\left(\omega_{1}z\right)}{2[1+\cos^{2}\left(\omega_{1}z\right)]}x^{2}+\frac{2\omega_{2}\cos(\omega_{2}z)}{1+\cos^{2}(\omega_{1}z)}+\gamma,
\end{equation}
 respectively, where 
\begin{eqnarray}
\gamma & = & -\frac{1}{4}G[1+\cos^{2}(\omega_{1}z)]^{2}\ln[(1+\cos^{2}(\omega_{1}z))/2]\nonumber \\
 &  & -\frac{4\omega_{2}^{2}\cos^{2}(\omega_{2}z)}{[1+\cos^{2}(\omega_{1}z)]^{2}}.
\end{eqnarray}

\begin{figure}[tb]
 \centering 
\includegraphics[width=0.48\columnwidth]{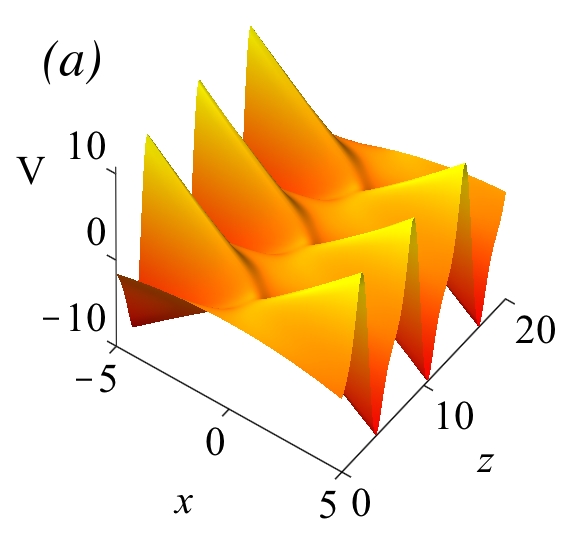}
\includegraphics[width=0.48\columnwidth]{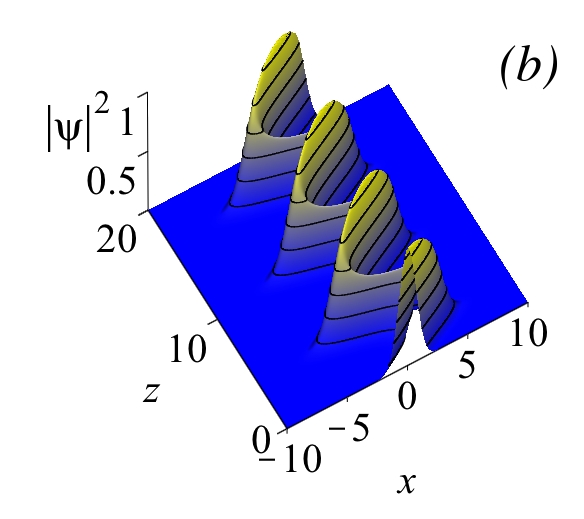} 
\caption{(Color online) (a) Linear coefficient (potential) and (b) modulated
solution for the case 3. We have used $G=-1$, $\epsilon=-G$, and
$\omega_{2}=2\omega_{1}=1$.}
\label{FC3} 
\end{figure}

In Fig. \ref{FC3} we depict the linear coefficient (potential) and
the profile of the solution ($|\psi|^{2}$) , considering $G=-1$,
$\epsilon=-G$, $\omega_{2}=2\omega_{1}=1$. This type of modulation
makes the solution to oscillate in the $x$-direction, with a breathing
profile. Also, solutions with quasiperiodic oscillation in $x$ and/or
z can be found with an appropriate adjustment of the ratio $\omega_{1}/\omega_{2}$
as an irrational number.

%%%%%%%%%%%%%%%%%%%%%%%%%%%%%%%%
%%%%%%%%%%%%%%%%%%%%%%%%%

%%%%%%%%%%%%%%
\emph{Stability analysis} - The numerical method is based on the split-step
Crank\textendash{}Nicholson algorithm in which the evolution equation
is splitted into several pieces (linear and nonlinear terms), which
are integrated separately. A given trial input solution is propagated
in time over small steps until a stable final solution is reached.
To this end, we have used the step sizes $\mathrm{\Delta x}=0.04$
and $\mathrm{\Delta z=0.001}$ that provide a good accuracy in the
final state \cite{MuruganandamCPC09}. To ensure the stability of
the method we also checked the norm (power) and the energy of the
solution defined by $P=\int_{-\infty}^{\infty}|\psi|^{2}\mathrm{dx}$
and 
\begin{equation}
E=\int_{-\infty}^{\infty}\mathrm{dx}\left\{ |\psi_{x}|^{2}+V|\psi|^{2}+g|\psi|^{2}\left(\log|\psi|^{2}-1\right)\right\} ,
\end{equation}
 respectively.

To study stability for the above cases we employ a random perturbation
in the amplitude of the solution with the form 
\begin{equation}
\psi=\psi_{0}[1+0.05\nu(x)],\label{pert}
\end{equation}
 where $\psi_{0}=\psi(x,0)$ is the analytical solution for the cases
1, 2, and 3, respectively, and $\nu\in[-0.5,0.5]$ is a random number
with zero mean evaluated at each point of discretization grid in $x$-coordinate.

\begin{figure}[tb]
 \centering
 \includegraphics[width=0.48\columnwidth]{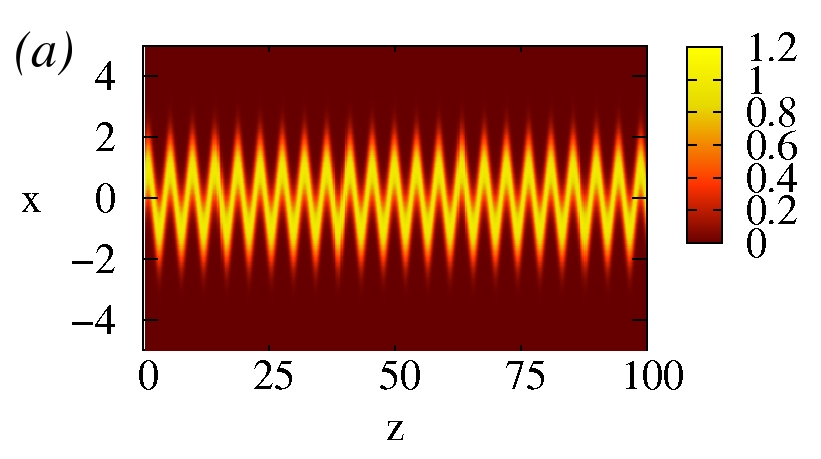} 
 \includegraphics[width=0.48\columnwidth]{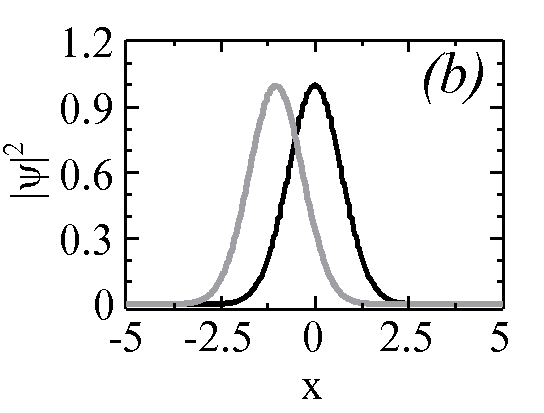}
\caption{(Color online) Plot of the solution profile for the case 1 (a) in
$x-z$ plane and (b) comparison between the input (black line) and
output (grey line) solution in the $x$ dimension. The input state
is taken at $z=0$ while the output state is in $z=1000$. Note in
(b) that the input and output present the same profile with different
peak positions due to the oscillations in the modulated snake-like
solution. This implies the stability of the solution, at least until
the observed $z$ value.}
\label{stab1} 
\end{figure}

In Fig. \ref{stab1} we show the numerical propagation of the input
state given by Eq.~(\ref{pert}) with $\psi(x,0)$ being the solution
of the case 1 and the comparison between the input ($z=0$) and output
($z=1000$) states. Note in Fig.~\ref{stab1}a that we have restricted
the profile to the value $z=100$ due to the large number of oscillations
when $z\gg100$. In this case the norm is maintained in $P\simeq1.76911$
with a standard deviation of $5\times10^{-13}$ while the energy oscillates
around $E\simeq90\pm14$.

%%%%%%%%%%%%%%%%%
%%%%%%%%%%%%%%%%%%%%%%%%%%%

\begin{figure}[tb]
 \centering 
 \includegraphics[width=0.48\columnwidth]{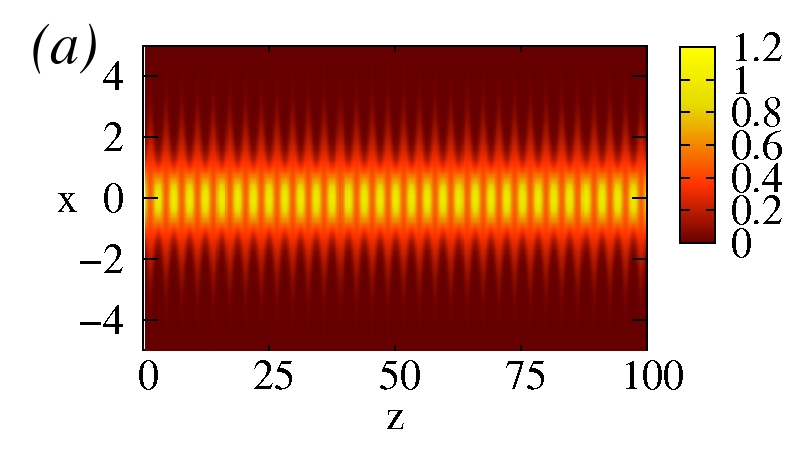} 
 \includegraphics[width=0.48\columnwidth]{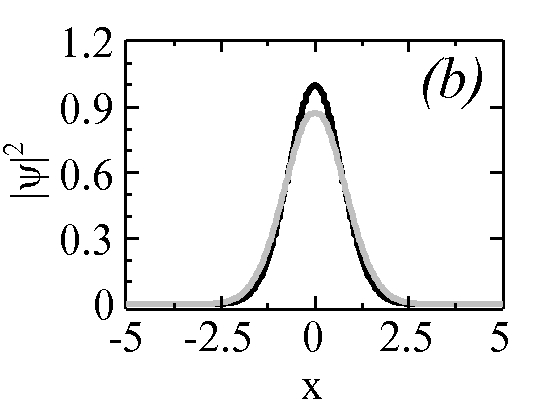}
\caption{(Color online) Plot of the solution profile for the case 2 (a) in
$x-z$ plane and (b) comparison between the input (black line) and
output (grey line) solution in the $x$ dimension. The input state
is taken at $z=0$ while the output state is at $z=1000$. Note in
(b) that the input and output present different peak amplitudes due
to the oscillations in the modulated solution.}
\label{stab2} 
\end{figure}

The numerical simulation of case 2 is displayed in Fig. \ref{stab2}.
Here the breathing pattern is preserved even when the input state
feels a small perturbation of the type shown in (\ref{pert}). Note
that the Fig. \ref{stab2}b presents a difference in the amplitude
of the input ($z=0$) and output ($z=1000$) states. This is due to
the oscillatory pattern of the solution, but we stress that it is
stable. We have obtained $P\simeq1.76911$ with a standard deviation
$\sim10^{-13}$ and with a respective energy given approximately by
$E\simeq49\pm12$ (with an oscillatory pattern).

%%%%%%%%%%%%%%%%%%%%%%%
%%%%%%%%%%%%%%%%%%%%%%%%%%%%%%

\begin{figure}[tb]
 \centering 
 \includegraphics[width=0.48\columnwidth]{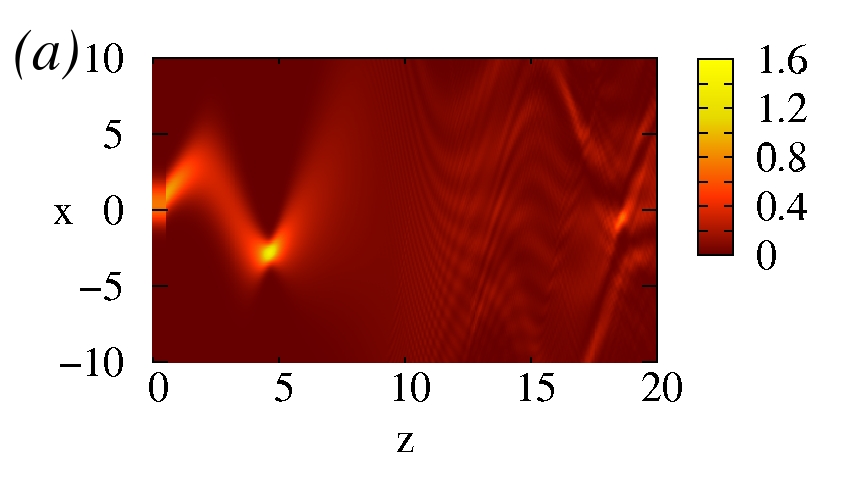} 
 \includegraphics[width=0.48\columnwidth]{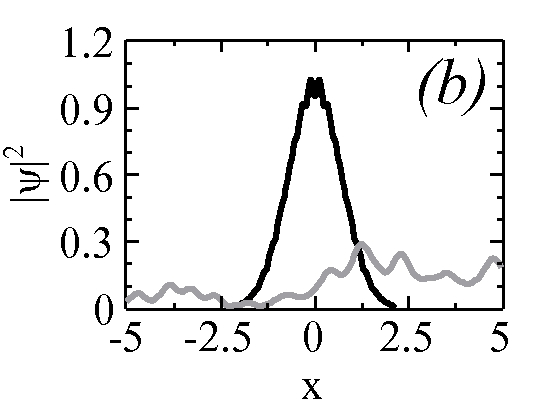}
\caption{(Color online) Plot of the solution profile (a) in $x-z$ plane and
(b) comparison between the input (black line) and output (grey line)
solution in the $x$ dimension for the case 3. The input state is
taken at $z=0$ while the output state is in $z=20$. Note that in
this case the modulation induces an unstable behavior.}
\label{stab3} 
\end{figure}

In the last simulation we have checked the instability for the case
3. In Fig. \ref{stab3} one can see the unstable behavior in the decay
of the solution. The norm is given by $P\simeq1.77096$ with a standard
deviation $\sim10^{-14}$ and the energy $E\simeq186\pm184$ (with
a random pattern due to the instability).

%%%%%%%%%%%%%%%%%%%%%%
%%%%%%%%%%%%%%%%%%%%%%%%

%%%%%%%%%%%%%%%%%%%
\emph{Conclusion} - In this work we investigated the presence of analytical
localized solutions to the LSE. We used similarity transformation
to deal with inhomogeneous nonlinearity and potential. The inhomogeneities
allowed us to modulate the pattern of the localized solution presenting
a snake-like, breathing, and mixed oscillatory and breathing forms.
The stability of the solutions was numerically checked and we have
shown some stable solutions for the model investigated.

\emph{Acknowledgement} - This work was supported by CAPES, CNPq, FAPESP
and INCT-IQ (Instituto Nacional de Ci\^encia e Tecnologia de Informa\c{c}\~ao
Qu\^antica).

%%%%%%%%%%%%%%%%%%%%%%%%

%%%%%%%%%%%%%%%%%%%%%%%%%%%%%%%%%%

\end{document}